\documentclass[10pt,a4paper,fleqn]{article}
\usepackage[utf8]{inputenc}
\usepackage{graphicx}
\usepackage{amsmath}
\usepackage{amssymb}
\usepackage[colorlinks]{hyperref}
\usepackage{authblk}

\def\ltop#1{\overline{t_{#1}}}
\def\lbot#1{\underline{t_{#1}}}

\newcommand\at[2]{\left.#1\right|_{#2}}
\begin{document}

\title{Moment-generating function of output stream of leaky integrate-and-fire neuron}
\author[1]{O.K. Vidybida}
\affil[1,2]{Bogolyubov Institute for Theoretical Physics of the National Academy of Sciences of Ukraine, Kyiv, Ukraine}
\affil[1]{vidybida@bitp.kiev.ua, vidybida.kiev.ua}
\author[2]{O.V.~Shchur}
\affil[2]{olha.shchur@bitp.kiev.ua}

\maketitle
\begin{abstract}
The statistics of the output activity of a neuron during its stimulation by the stream of input impulses that forms the stochastic Poisson process is studied.
The leaky integrate-and-fire neuron is considered as a neuron model.
A new representation of the probability distribution function of the output interspike interval durations is found. Based on it, the moment-generating function of the probability distribution is  calculated explicitly. The latter, according to Curtiss theorem,
completely determines the distribution itself. In particular,  explicit expressions are derived from the moment-generating function for the moments of all orders. The first moment coincides with the one found earlier. Formulas for the second and third moments have been checked
numerically by direct modeling of the stochastic dynamics of a neuron with specific physical parameters.
\end{abstract}


\section{Introduction}
\begin{figure}
\centering
\includegraphics[width=\textwidth]{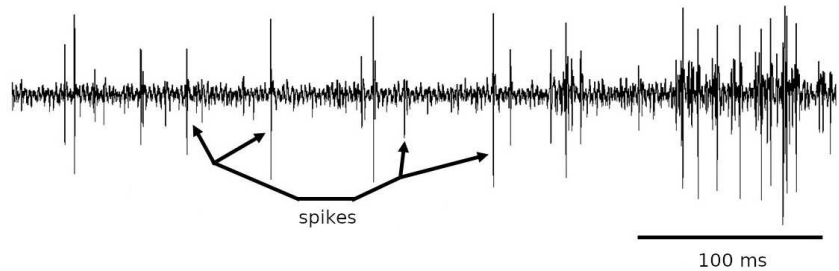}
\caption{\label{spikes}
An example of electrical activity in the environment of neurons.
The recording is made by an electrode, which is placed in the environment. Therefore, the activity of several neurons is registered simultaneously.
In a case of a neuron generating an output impulse, its membrane potential changes drastically.
At these moments, the figure  shows short-term jumps of voltage, or spikes.
Different spikes have different heights because they belong to different neurons that are at different distances from the recording electrode. A single neuron generates spikes of the same height.
 Modified from: {\tt https://backyardbrains.com/experiments/spikerbox}.
}
\end{figure}

Information in the brain is mainly represented in the form of neural impulses.
All those impulses are roughly identical in their height and width and called
spikes, see Fig. \ref{spikes}.
The only thing which matters is the time when such an impulse
has been generated or received.
If neural impulses are recorded with proper biophysical instruments, one obtains a highly
irregular sequence. It is called a spike train.
It is difficult to find any rational meaning in these moments
of receiving spikes, or interspike
intervals. The situation is even worse: In most cases, those sequences do not reproduce
themselves if the same stimulation is offered to an experimental animal
several times.
This might be the first reason why neuroscientists are mainly interested
in the statistical properties of spike trains.
Theoretical physicists as well try to predict which kind of statistics it could
be and how does it change with changing stimuli or model parameters.
In this direction, there is a long-standing discussion.
What indeed represents meaningful information in a spike train?
Is it the mean number of spikes per time unit (rate coding), or are their exact temporal
positions (time coding) essential?
There is no clear answer to this question.
Initially, we supposed that the rate coding operates at the periphery of
the nervous system. An example at the actuator periphery of the brain is
the neuro-muscular junction in motoneurons \cite[Sec. 5.01.12]{Werner2014}. Namely, 
the only command passed to the muscle from the motoneuron is the 
contraction strength. But the contraction strength
is determined by the neurotransmitter concentration, which is released from
neural endings with each spike arrival. The more spikes per time unit,
the higher the neurotransmitter level, the higher the contraction
strength. So, here we have the rate coding.
An example at the sensory periphery of the brain is the olfactory receptor neuron \cite{Brookes2011},
where the number of spikes per time unit  depends on the odor concentration.

However, even at the sensory periphery, the time coding can be the coding paradigm.
This is observed for echolocation \cite{Denny2004}, where the temporal position of spikes
from two ears should be kept with microsecond precision.

It is also clear that in the time coding mechanism a spike train can bear
more information than in the rate coding one. This could be essential for more
sophisticated intellectual tasks than muscle contraction or odor sniffing.

\begin{figure*}
\centering \includegraphics[width=0.7\textwidth]{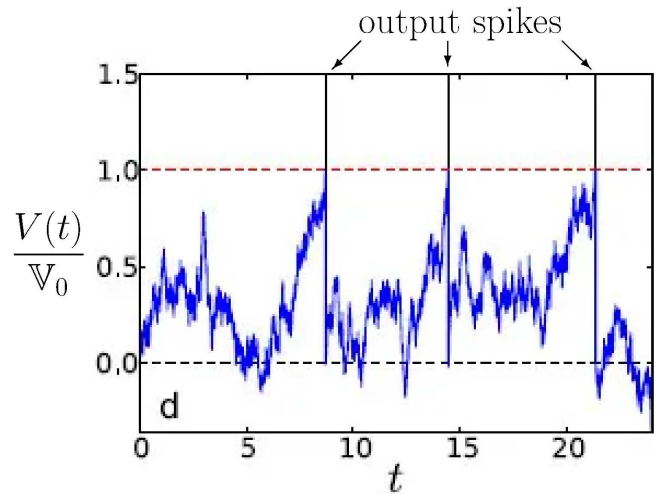}
\caption{\label{finite}The time course of the membrane potential, simulated by the diffusion process. At the moment when the membrane potential $ V (t) $ reaches the firing threshold $ \mathbb V_0 $, the output impulse (spike) is generated.
 Modified from:     A Iolov, S Ditlevsen \& A Longtin, DOI 10.1186/2190-8567-4-4.
 }
\end{figure*}

Unfortunately, most attempts to calculate neuronal firing statistics exclude the
possibility of time coding due to utilizing the so-called diffusion
approximation.
In this approach, the neuronal stimulus is modeled as a diffusion stochastic process
such as Wiener or Ornstein-Uhlenbeck one, see \cite{Gardiner1985,Sacerdote2013}. 
In a diffusion process, any finite
time interval contains infinitely many infinitesimal spikes obtained
from the differentiation of a Wiener process.
Therefore, there is no place for the time coding mechanism.
At the same time, the output activity of a neuron stimulated with a
diffusion process is represented by finite spikes emitted when neuronal
membrane voltage crosses the firing threshold. 
The time intervals between those spikes are finite, see Fig. \ref{finite}.
Those spikes represent not diffusion but point process and, therefore, cannot be
fed into another neuron preserving the diffusion approximation approach.
That means that the diffusion approximation approach is both incomplete and
inconsistent.
Therefore, an attempt was made to calculate firing statistics without
diffusion approximation.

In the following sections, we will briefly formulate previously obtained results this work is based on. Preliminary results are related to the statistics of the activity of the leaky integrate-and-fire (LIF) neuron with a threshold of 2. In particular, earlier, \cite{Vidybida2014a, Vidybida2016}, explicit expressions were obtained for the distribution function of the output interspike intervals (ISI) at the initial section of the values of the ISI duration. The distribution of input impulses is considered to be Poissonian.
For larger values of ISI, the distribution function is represented as the sum of multiple integrals.
This enabled us to calculate the first moment of the distribution function (mean ISI).
In the current paper, we have found another representation of the distribution function, which provides the means to calculate the moment-generating function. Applying differentiation to it, a distribution moment of any order can be found.
Also, according to Curtiss theorem \cite{Curtiss1942}, the moment-generating function completely determines the distribution function itself.

\section{Preliminary results}

\subsection{Model description}\label{model}

 The leaky integrate-and-fire neuron \cite{Burkitt2006}  is characterized by three positive constants:
 $\tau$ is the relaxation time;
 $\mathbb V_0$ is the firing threshold;
 $h$ is the input impulse height.
 
 At any given time $t$ the state of LIF neuron is defined by the non-negative
real number $V(t)$, which is the deviation of the transmembrane potential difference from the rest state towards depolarization, or, in other words, the magnitude of the excitation. Here it is assumed that at rest $ V = 0 $, and the depolarization corresponds to a positive value of $V$.
The presence of the leakage means that in the absence of external stimuli, the value of $ V (t) $ decreases exponentially:
\begin{equation}\label{tau}
	V(t+s)=V(t) e^{-s/\tau},
	\qquad s>0.
\end{equation}
Input stimuli are input impulses.
An input impulse obtained at time $ t $
increases $ V (t) $ by $ h $:
\begin{equation}
\label{h}
	V(t) \rightarrow V(t) + h.
\end{equation}
The neuron is characterized by the firing threshold $ \mathbb V_0 $.
The latter means that once the condition is met, i.e.
$
	V(t) > \mathbb V_0,
$
\ LIF neuron generates an output impulse
and resets to the rest state, $ V (t) = 0 $.
Regarding $ h $ and $ \mathbb V_0 $, we make the following assumption: 
\begin{equation}\label{cond}
	0<h< \mathbb V_0 < 2h.
\end{equation}
From (\ref{tau}) and (\ref{h}), it follows that the LIF neuron can generate an output impulse only at the time of receiving the input one.
The condition (\ref{cond}) means that one input impulse,
applied to the LIF neuron at rest, is not enough to generate the output impulse. However, even two input impulses obtained in a short time can excite the LIF neuron above the threshold and generate the output impulse. This means that the neuron has a threshold of 2.\footnote{See \cite{Kravchuk2016}, where the case of higher thresholds is considered.}

\subsection{Distribution function of ISI durations\protect\footnote{In the current section and below, $t$ denotes  ISI duration.}}

We assume that described in Sec. 2.1 neuron is stimulated by the stream of input impulses, which forms a stochastic Poisson process of intensity $ \lambda $.
The latter means that the probability of obtaining  the ISI of duration $ t $ with the precision $ dt $ at the input is given by the following expression:
\begin{equation} e^{-\lambda t}\, \lambda\,dt,
\end{equation}
and input ISIs are statistically independent.

We introduce the following notations:
\begin{multline}\label{T23}
T_2 = \tau\ln(\frac{h}{\mathbb V_0-h}),~ T_3 = \tau\ln(\frac{\mathbb V_0}{\mathbb V_0-h}),~
\\
\Theta_m=T_2 + (m-3)T_3,~ m=3,\dots.
\end{multline}

In papers \cite{Vidybida2014a,Vidybida2016},
the following formula is obtained for distribution function of output ISI: 
\begin{multline}\label{Sumiii}
P(t)dt= 
\sum\limits_{k=2}^{m-1} 
\left(P_k^0(t)\lambda dt - P_k^-(t)\lambda dt\right)
+P_m^0(t)\lambda dt,
\\
t\in\,]\Theta_m;\Theta_{m+1}],\, m\ge2 ,
\end{multline}
where $P(t)dt$ is the probability to obtain output ISI of duration $ t $ with the precision $ dt $.
Functions on the right side of (\ref{Sumiii}) are defined as follows:
\begin{multline}\label{F-la}
 P_{k+1}^0(t)\lambda\,dt=
 \int\limits_{\Theta_{k+1}}^t P_k^-(s)\lambda\, ds\,e^{-\lambda(t-s)}\lambda\,dt,
\\
 t\ge \Theta_{k+1},\quad
k=2,3,\dots ,
\end{multline}
\begin{equation}\label{P23}
P_k^-(t)\lambda dt = e^{-\lambda t}\lambda^k dt 
\int\limits_{\lbot1}^{\ltop1} dt_1
\int\limits_{\lbot2}^{\ltop2} dt_2
\dots
\int\limits_{\lbot{k-1}}^{\ltop{k-1}} dt_{k-1},
\end{equation}
where the limits of integration are defined through the following inequalities:
\begin{equation}\label{P23limits}
\begin{cases}
0 \le t_1 \le t-\Theta_{k+1},
\\
  T_2 +
\tau\ln\left(\sum\limits_{1\le j\le i} e^{t_j/\tau}\right)
\le
 t_{i+1},
\\
 t_{i+1}
\le
 \tau\ln\left(e^{(t-\Theta_{k+1-i})/\tau}-\sum\limits_{1\le j\le i} e^{t_j/\tau}\right),
 \\
 i=1,\dots,k-2.
\end{cases}
\end{equation}
Thus, the distribution function of the output ISI is completely determined by the function $P_k^-(t)$ for different $ k = 2,3, \dots $.
Its physical meaning is as follows: if the neuron starts from the rest state, $ V(0) = 0 $, then the expression $ P_k^-(t) \lambda dt $ gives the probability to obtain from the Poisson input process $ k $ consecutive input impulses in such a way that the last of them falls into the interval
$ [t; t + dt [$ and the neuron does not fire (the excitation threshold $ \mathbb V_0 $ has not been exceeded). In turn, $ P_{k}^0 (t) \lambda \, dt $ gives the probability to obtain $ k $ impulses,
the last one within the interval $ [t; t + dt [$ so that there is no firings up to and including the (k-1)-th impulse.
Note that in the formula (\ref{P23}) for a fixed $ t $, $ k $ cannot take values greater than $ k_{max} $, where
\begin{equation}\label{kmax}\nonumber
k_{max}=\left[\frac{t-T_2}{T_3}\right]+2,
\end{equation}
and square brackets denote an integer part of a number.

\section{A new representation of the distribution function}
In this section we will represent (\ref{P23}), (\ref{P23limits}) in a simpler form, convenient for calculating the moment-generating function. For this reason, we introduce new integration variables:
\begin{equation}\label{newz}\nonumber
z_i=e^{-\frac{t-\Theta_{k+2-i}}{\tau}}\sum\limits_{1\le j\le i}e^{\frac{t_j}{\tau}},\quad i=1,\dots,k-1.
\end{equation}
The domain of integration (\ref{P23limits}) in terms of new variables takes the following form:
\begin{equation}\label{P23limitszz}
\begin{cases}
e^{-\frac{t-\Theta_{k+1}}{\tau}} \le z_1 \le  1,
\\
z_i \le z_{i+1}\le 1,\quad i=1,\dots,k-2.
\end{cases}
\end{equation}
The Jacobian determinant of the transition to the new variables has the following form:
\begin{equation}
\left|\det||\frac{\partial z_j}{\partial t_i}||\right|=
\frac{1}{\tau^{k-1}}z_1\prod\limits_{2\le i\le k-1}(z_{i-1} - \beta z_i),
\end{equation}
where $\beta=e^{-\frac{T_3}{\tau}}$.
Taking into account (\ref{P23limitszz}) and the latter, (\ref{P23}) can be expressed in the following form:
\begin{equation}\label{P23new}
P_k^-(t) = e^{-\lambda t}(\lambda\tau)^{k-1} 
\int\limits_{B_k(t)}^{1} \frac{dz_1}{z_1}
\int\limits_{z_1}^{1} \frac{dz_2}{z_2-\beta z_1}
\dots
\int\limits_{z_{k-2}}^{1} \frac{dz_{k-1}}{z_{k-1}-\beta z_{k-2}},
\end{equation}
where
\begin{equation}\label{B(t)}\nonumber
B_k(t)=e^{-\frac{t-\Theta_{k+1}}{\tau}}.
\end{equation}\bigskip
If the set of auxiliary functions  $f_i(x)$  is introduced by the following
relations:
\begin{equation}\label{fdef}
f_0(x)\equiv 1,~ f_{i+1}(x) = \int\limits_{x}^{1} \frac{dy}{y-\beta x}f_i(y),~
i=0,\dots,
\end{equation}
then (\ref{P23new}) can be written as
\begin{equation}\label{P23neww}
P_k^-(t) = e^{-\lambda t}(\lambda\tau)^{k-1} 
\int\limits_{B_k(t)}^{1} \frac{dx}{x}
f_{k-2}(x).
\end{equation}
The latter with (\ref{Sumiii}) and (\ref{F-la}) are used below to calculate
the moment-generating function.

\section{Moment-generating function}
The moments of probability distribution $P(t)$ are the quantities $\mu_n$ given by the formula\footnote{In considered case, $t\le0 \Rightarrow P(t)=0$.}
\begin{equation}\label{momentsDef}
\mu_n= \int_{-\infty}^\infty t^n P(t)dt.
\end{equation}
Here the first moment is the mean of a random variable, in our case of ISI. Moments calculation of  can be difficult given the complexity of the expression for $ P (t) $. The moment-generating function  simplifies the task.

According to the definition, the moment-generating function $ M_t (z) $ is determined by the following formula:
\begin{equation}\label{defM}
M_t(z) = \mathbb{E}[e ^{tz}] = \int_{-\infty}^\infty e ^{tz} P(t)dt.
\end{equation}

To find it, let us represent the distribution function $ P (t) $ (\ref{Sumiii}) through
auxiliary functions $ f_i (x) $ (\ref{fdef}). To accomplish this, firstly,  the expression (\ref{F-la}) for $ P_ {k}^0 (t) $ should be rewritten through $ f_i (x) $ (\ref{fdef}), substituting there (\ref {P23neww}): 
\begin{equation}\label{p+}
P^0_{k+1}(t)=
\lambda(t-\Theta_{k+1})P^{-}_k (t)+e^{- \lambda t} r^{k}\int_{B_{k+1}(t)}^1 \dfrac{\ln(x)}{x}  f_{k-2}(x) dx,
\end{equation}
where $r=\lambda \tau$. 

By regrouping the terms in the sum on the right-hand side (\ref{Sumiii}) and substituting there (\ref{P23neww}) and (\ref{p+}), the following expression for the distribution function $ P (t) $ can be obtained through the functions $ f_i (x) $:
\begin{multline}\nonumber
P(t)dt = 
\lambda t e^{- \lambda t}dt+e^{- \lambda t}dt
\sum_{k=3}^{m}r^{k-2}\int_{B_{k}(t)}^1 \dfrac{dx}{x}f_{k-3}(x) (\lambda(t-\Theta_{k})-1+r \ln(x)),\\
t\in ]\Theta_m;\Theta_{m+1}],\:m\geq 2.
\end{multline}

The latter is used in (\ref{defM})  to find the moment-generating function:
\begin{equation}\label{M}
M_t(z)=
\dfrac{\lambda^2}{(\lambda-z)^2}+\dfrac{\lambda z}{(\lambda-z)^2}\sum_{m=3}^{\infty} r^{m-2} e^{-( \lambda-z)\Theta_m} I_m (z),\: z<\lambda,
\end{equation}
where the following auxiliary functions $I_m(z)$ are introduced:
\begin{equation}\label{Im}\nonumber
I_m(z)=\int_0^1 dx\; f_m(x) x^{r-\tau z-1},\;m=0,1,\ldots.
\end{equation}

To find recurrent relation for $I_m(z)$, substitute (\ref{fdef}) into the latter: 
\begin{equation}
\label{reccurent}
I_m(z)=
\Phi (\beta,1,r-\tau z) I_{m-1}(z),\;m=1,2\ldots;\;
I_0(z)=\dfrac{1}{r-\tau z}.
\end{equation}

Here $\Phi (\beta,1,r-\tau z)$  denotes Lerch transcedent:
\begin{equation}\nonumber
\Phi (z,s,a) = \dfrac{1}{\Gamma(s)}\int_0^1 \dfrac{dx}{1-z x} (-\ln (x))^{s-1} x^{a-1}.
\end{equation}

It follows from the reccurent relation (\ref{reccurent}) that
 \begin{equation}\begin{split}\label{I}\nonumber
I_m(z)= \dfrac{1}{r-\tau z} \left(\Phi \left(\beta,1,r-\tau z\right)\right)^m ,\;m=0,1\ldots.
\end{split}\end{equation}

Substitute the latter into (\ref{M}) and use the definition of $\Theta_m$ (\ref{T23}):
\begin{multline}\label{MwithSum}
M_t(z)
 =\dfrac{\lambda^2}{(\lambda-z)^2}+
 \dfrac{\lambda z}{(\lambda-z)^2} \dfrac{r}{r-\tau z} e^{-( \lambda-z)T_2}\times\\
\times \sum_{m=0}^{\infty}
 \left(r \beta^{r(1-\frac{z}{\lambda})} \Phi \left(\beta,1,r-\tau z\right)\right)^{m}.
\end{multline}

Here the series $\sum_{m=0}^{\infty}
 \left(r \beta^{r(1-\frac{z}{\lambda})} \Phi \left(\beta,1,r-\tau z\right)\right)^{m}$ is convergent in some neighbourhood of the point $z=0$, since $
r \beta^{r} \Phi \left(\beta,1,r \right)<1.$ The latter is proved in the Theorem 3 of
\cite{Vidybida2016}.

Finally, after summing in the right-hand side of (\ref{MwithSum}), in some neighbourhood of the point $ z = 0 $, the moment-generating function has the following form:
\begin{equation}\label{tvirna}
M_t(z)=
 \dfrac{\lambda^2}{(\lambda-z)^2}+ a^r
 \dfrac{\lambda z}{(\lambda-z)^2} \dfrac{r}{r-\tau z} 
  \dfrac{e^{z T_2}}{1-r \beta^r e^{zT_3} \Phi (\beta,1,r-\tau z)},
\end{equation}
where $a=e^{-\frac{T_2}{\tau}}$.

Since in some neighborhood of zero the moment-generating function is finite, then, according to Curtiss theorem \cite{Curtiss1942}, the obtained moment-generating function (\ref{tvirna}) completely determines the distribution function $ P (t) $.
 
Using the moment generating function (\ref{tvirna}), the moments of the distribution function can be found:
\begin{multline}\label{mn}
\mu_n = \at{\dfrac{d^n M_t(z)}{d z^n}}{z=0}=
\dfrac{(n+1)!}{\lambda^n}+ \dfrac{n!a^r }{2\lambda^n}\dfrac{1}{1-r \beta^r  \Phi (\beta,1,r)}\times \\
\times \sum_{m=0} ^{n-1}\dfrac{(\lambda(T_2-T_3))^{m}}{m!}  \sum_{k=0} ^{n-1-m} (n-m-k)(n-m-k+1)\times \\
\times \Bigg( \delta_{k,0}
 +\dfrac{1}{k!}\sum_{l=1}^{k}
\dfrac{(-1)^l l!}{(1-r \beta^r  \Phi (\beta,1,r))^{l}}B_{k,l} (g_1,g_2,\ldots,g_{k-l+1})\Bigg),\\
g_m=(-\lambda T_3)^m-m!r^{m+1} \beta^r \Phi(\beta, m+1, r);
\end{multline}
where $\mu_n$ denotes the $n$-th moment, $B_{k,l} (g_1,g_2,\ldots,g_{k-l+1})$ are
incomplete exponential Bell polynomials.

Setting in the last expression $ n = 1 $, for the first moment, we have:
\begin{equation}\nonumber
\begin{split}
\mu_1 = 
\dfrac{2}{\lambda}+\dfrac{1}{\lambda} \dfrac{a^r}{1-r\beta^r \Phi(\beta,1,r)},
\end{split}
\end{equation}
which coincides with obtained previously in
 \cite[Eq. (46)]{Vidybida2016}. Notice that in terms used in work \cite{Vidybida2016} $ I(a,r)=\beta^r \Phi(\beta,1,r)$.

According to (\ref{mn}) in case of  $n=2$, the second moment is the following:
\begin{multline}\label{m2}
\mu_2 =\dfrac{6}{{\lambda}^2}+\dfrac{2}{\lambda^2}\dfrac{a^r}{1-r \beta^r \Phi(\beta,1,r)}\Bigg(3+\lambda T_2 +\\
+ \dfrac{r \beta^r \Phi(\beta,1,r)}{1-r \beta^r \Phi(\beta,1,r)}
\left(\lambda T_3+r \dfrac{\Phi(\beta,2,r)}{\Phi(\beta,1,r)} \right) \Bigg).
\end{multline}

\section{Numerical verification}
\begin{figure*}
\centering \includegraphics[width=\textwidth]{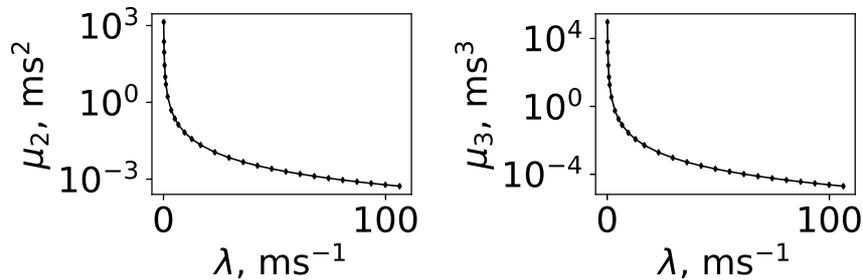}
\caption{\label{moments} Dependence of the second $ \mu_2 $ and the third $ \mu_3 $ moments of the ISI duration distribution function on the intensity of the input stream $ \lambda $. Diamonds denote results of numerical simulation by the Monte Carlo method, solid line --- calculations according to the formulas (\ref{m2}) and (\ref{mn}) at $ n = 3 $. Here $\mathbb V_0=20$ mV, $h=11.2$ mV, $\tau=20$ ms.
 }
\end{figure*}

To numerically verify the obtained formulas, a program was written that simulated the dynamics of the membrane potential of a neuron stimulated by the stream of input impulses that form the stochastic Poisson process. The behavior of the neuron was simulated for such a time that, as a result, 1 000 000 output impulses were obtained, which allowed calculation of the probability density $ P (t) $ and its moments, as shown in (\ref{momentsDef}). The simulation was repeated for different values of the input stream intensities $ \lambda $. The results of calculating the 2nd and 3rd moments and their comparison with the formulas (\ref{m2}) and (\ref{mn}) for $ n = 3 $ are shown in Fig. \ref{moments}.

\section{Conclusions}
In the current paper, the statistics of the activity of the leaky integrate-and-fire neuron during its stimulation by input impulses, which form the stochastic Poisson process, is considered. For the model of a neuron with a threshold of two, a comprehensive description of the statistics of the durations of interspike intervals  in terms of the moment-generating function  is obtained. The latter is found explicitly, Eq. (\ref{tvirna}). The obtained formulas have been verified by numerical modeling of neuron dynamics with specific physical parameters.
\bigskip

{\small
Acknowledgments. This work was supported by the Program of Basic Research of the Department of Physics and Astronomy of the National Academy of Sciences of Ukraine ``Noise-induced dynamics and correlations in nonequilibrium systems'',
№ 0120U101347.
}
\bibliographystyle{unsrt}


\end{document}